\documentclass[%
reprint,
showpacs,
 amsmath,amssymb,
 aps,
 pre,
]{revtex4-1}
\usepackage{graphicx}
\usepackage{dcolumn}
\usepackage{bm}
\begin{document}

\title{Mesoscopic Simulation of Electrohydrodynamic Patterns in
Positive and Negative Nematic Liquid Crystals}
\author{Kuang-Wu Lee}
\email{jeff.lee@fau.de}

\author{Thorsten P\"oschel}%
\affiliation{Institute for Multiscale Simulation, Erlangen-Nuremburg University, Erlangen, Germany}
\date{\today}
\begin{abstract}
For the first time the electrohydrodynamic convection (EHC) of nematic liquid crystals is studied
via fully nonlinear simulation. As a system of rich pattern-formation the EHC is mostly studied with negative nematic
liquid crystals experimentally, and sometimes with the help of theoretical instability analysis in the linear regime.
Up to now there is only weakly nonlinear simulation for a step beyond the emergence of steady convection rolls.
In this work we modify the liquid crystal stochastic rotational model (LC SRD)
[Lee \textit{et al., J. Chem. Phys.}, 2015, \textbf{142}, 164110] to incorporate
the field alignment mechanism for positive and negative nematic liquid crystals.
The convection patterns and their flow dynamics in the presence of external electric field are studied.
Our results predict the similar optical convection patterns in the reflected polarized light when one uses
positive and negative types of nematic liquid crystals. However in the emerged flow fields, surprisingly,
the driving areas of convection rolls are different for different types of LCs. Their application for nematic
colloidal transportation in microfluidics is discussed.

\end{abstract}

\maketitle

\section{Introduction}

Electrohydrodynamic convection \cite{Roberts1968} in nematic liquid crystals is an interesting topic
since its discovery in 1963 by Williams \cite{Williams1963}. It attracts great amount attention due
to the rich pattern formation phenomena and profound potentials in industrial use.
Similar to Rayleigh-B\'enard convection (RBC) in simple fluids EHC is a demonstration of hydrodynamic instability
resulted from competition between external driving force and the internal dissipation force. The scientific
interests on EHC and RBC are mainly addressed on the phase transitions from homogeneous state to regular and turbulent
patterns.

From liquid crystal industrial point of view, EHC related instabilities are of fundamental
importance because most of the liquid crystal displays (LCDs) are driven by external electric field. Because
the DC electric field degrades the LC molecular structure in time, AC electric field ranging in $60 - 600Hz$ is
mostly used for display usage. In this frequency range a low driving voltage is enough for EHC instability,
which is usually accompanied with topological defects that destroy the optical properties of LCDs.
On the other hand a low driving voltage is desirable for low consumption, therefore the searching for stable
operational frequency range without turbulent EHC pattern is one of the central focuses.

With the potential usage of guided colloidal transport, Sasaki et al. \cite{Sasaki2014} created a
caterpillar type of nematic colloidal chain that could be transported by the EHC in the microfluidic channel.
The guided colloidal chain was also demonstrated to carry a silicone oil droplet and a glass rod as loaded microcargos in the channel. The controlled transport processes are important for, e.g., drug delivery,
lab-on-a-chip applications, and directed self-assembly. EHC provided a new method than by using the known
active entities, such as bio-molecular motors, synthetic nano-machines, and by external stimuli.

To understand the generation of EHC in liquid crystals, a driving mechanism has been proposed by 
Carr \cite{Carr1969}. Linear perturbation theory was subsequently applied to the EHC system to study the
onset of the instabilities \cite{Orsay1970}. Beyond the initial onset of EHC, weakly nonlinear theory considering
mean flow effects \cite{Bodenschatz1988} is used to study the evolution from ordered periodic to weakly turbulent
patterns. As in many systems that undergo phase transition, it is observed that the generation of EHC rolls is due
to the thermal noise of the system \cite{Rohberg1991}. In their work the system intrinsic thermal noise corresponds
to the director fluctuation at the EHC onsets, indicates the external electric field is feeding energy to the most
unstable mode therefore causing the instability growth.

Despite the abundant exciting experimental works and the weakly nonlinear theory for the onset instability, numerical
investigations for the full EHC development still fall behind. EHC is fundamentally a highly nonlinear phenomena
therefore a numerical investigation is necessary for its evolution. To analyze this complex reaction-diffusion
process the first weakly nonlinear simulations \cite{Rossberg1989} was performed by neglecting the high order
perturbation terms in the hydrodynamic equations,
this approach was used to study the transition from normal to oblique EHC rolls. The optical properties of EHC
patterns in thin cells are studied in finite-difference-time-domain (FDTD) method \cite{Bohley2005}, for which
the reflective and transmitted lights of normal EHC rolls are derived by assuming background director and flow
fields. The dynamics of topological defects generated by positive and negative type of LCs is simulated by a
lattice model using Lebwohl-Lasher potential \cite{deOliveira2010}. External electric field is applied and it
is found that the field
can remove certain type of defect charges according to the LCs used. Nevertheless the coupling of director and
flow fields is neglected in their 2D model for simplicity reason, a fully coupled dynamics should be restored
to describe the dynamics precisely. Therefore our summary of the current status of EHC numerical study should be
concluded that a fully nonlinear simulation of a self-consistent EHC evolution is still not provided.

In the present study we introduce the particle-based stochastic rotation dynamics (SRD) model for the
nonlinear evolution of EHC. This model is recently developed by Lee et al. \cite{Lee2015} for the study
of topological defects in nematic liquid crystals. It is proved in that work the thermal fluctuation
determines the nematic-isotropic phase transition in NLC, for which similarly the thermal fluctuation
controls the evolution of EHC patterns \cite{Rohberg1991}.
Modifications on the molecular potential and the momentum equation could couple the LC rotation to the
external electric field therefore enable us to study EHC nonlinearly. Positive and negative types of
liquid crystals, such as 5CB and MBBA, are modeled by changing the anisotropy of dielectric tensor
$\epsilon_a = \epsilon_{\parallel} - \epsilon_{\perp}$.
The EHC flow patterns are revealed in this fully nonlinear simulations, and the optical properties of
the reflected lights are compared accordingly. We will address our attention on the similarity and
distinction of its flow and reflection fields, when different type of liquid crystals are used.

\section{Model}

Stochastic rotation dynamics \cite{Malevanets1999} (SRD) is a particle-based algorithm consists two steps, i.e.
the free streaming step for updating the particle positions and the rotational step to mimic the velocity
change during particle collisions. For anisotropic fluids such as liquid crystals, one extra degree of freedom
is added to the particle orientation, indicating the particle director $\vec{d_i}$. Liquid crystal is a material
of strong coupling between the fluid velocity and the director field, therefore the governing equations of
fluid velocity and director orientation should have feedback from each others. The validation of liquid crystal
model using SRD was proposed by Lee et al. \cite{Lee2015}, and nearly simultaneously a slightly different but
independently developed model was proposed by Shendruk and Yeomans \cite{Shendruk2015}. Several important LC
phenomena are successfully observed this particle-based mesoscopic model, e.g. the first-order nematic-isotropic
phase transition, the dynamics of topological defects and the non-newtonian shear-banding effect.
\begin{figure}[h]
\centering
  \includegraphics[height=3cm]{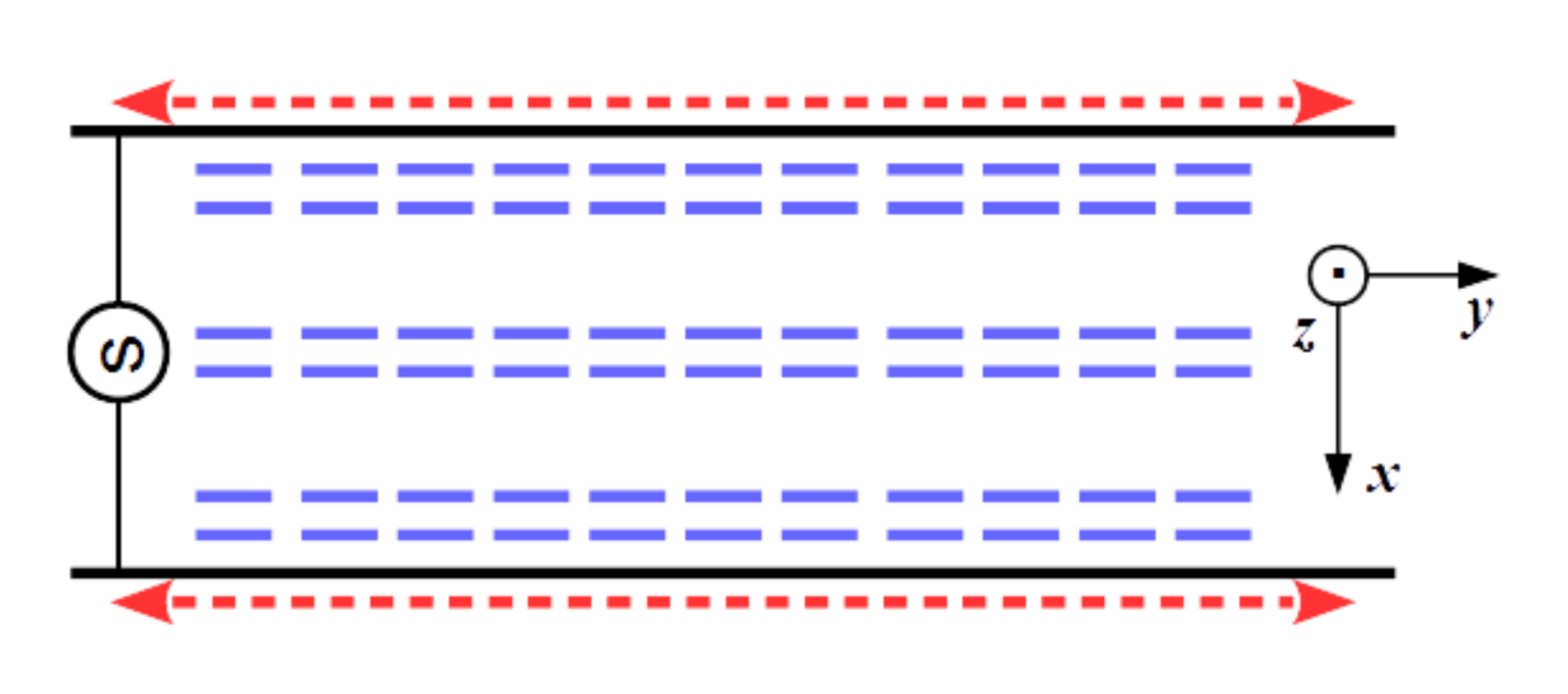}
  \caption{The configuration of simulation setup. The red dashed line indicates the wall anchoring in
  y direction, and the blue line segments are the nematic LCs. AC electric field is along x direction $\vec E_x$.}
  \label{system}
\end{figure}
The advantage of a particle-based fully nonlinear simulation over nematohydrodynamic approach is that, the unstable
modes are generated by the particle thermal noise, i.e. it avoids 1) the phase-space scanning of initial unstable
wave modes and 2) assuming non-zero amplitude of those eigenmodes. Those process required by weakly nonlinear analysis
and simulation are self-consistently generated by stochastic particle motion.

Consider the external electric field, the nematohydrodynamic system we would like to simulate is the simplified
Ericksen-Leslie model \cite{Lin1995}. The governing equations for the cell-wise bulk flow $\vec v$ and director
$\vec d$ are as follow \cite{Breindl2005,Lee2015}

\begin{gather}
\frac{\partial\vec v}{\partial t}+ \vec v\cdot \nabla \vec v =\nabla\cdot(\nu\nabla\vec v)-\nabla P/\rho
-\lambda\nabla \cdot \pi + \pi^2 \rho_e \vec E
\label{Navier-Stokes}
\\
\frac{\partial\vec d}{\partial t}+ \vec v\cdot \nabla \vec d
- \vec d \cdot \nabla\vec v = \gamma_{_\mathrm{EL}} \nabla^2 \vec d - \gamma f(\vec d) + \vec \xi(t)
\label{EL-director}
\end{gather}

In Eq.(\ref{Navier-Stokes}) the charge density $\rho_e = \nabla \cdot (\epsilon \vec E)$ can be obtained
from Poisson's equation, where the dielectric tensor is
$\epsilon_{i,j} = \epsilon_{\perp}\delta_{i,j} + \epsilon_{a}d_i d_j$.
In Eq.(\ref{EL-director}) the molecular field $f(\vec d_{i}) = \partial U_{i}/ \partial \vec d_{i}$
is the vector derivative of the molecular potential. In the presence of external electric field
$\vec E$ a modified Lebwohl-Lasher type \cite{Lebwohl1972} molecular potential is used.

\begin{equation}
U_{i} = - \epsilon_{a} (\vec d_{i} \cdot \vec E_{ext})^2
-\sum_{\left\langle i,j\right\rangle} (\vec d_{i} \cdot \vec d_{j})^2
\label{LL-potential}
\end{equation}

Where the indices $i$ and $j$ are the particle considered and the particles around it. The external electric 
field $E_{ext} = -dV/ds$ is expressed as voltage difference across one cell $dV$ in our simulations. 

The system to be simulated here is only slightly different from the model used in Lee et al. \cite{Lee2015},
for which the angular momentum change due to the field-induced LC rotations is further balanced in the
Navier-Stokes equation Eq.(\ref{Navier-Stokes}). For the conservations of linear momentum and kinetic energy
the SRD algorithm can achieve in the collision step. Because of the extra degree of freedom of LC oreintation,
the Lebwohl-Lasher model is used to describe the conservation of angular momentum. One can easily prove that
the last two terms in the Eq.(\ref{Navier-Stokes}) are for the balancing of angular momentum change due to the
LC flow rheology and the external electric field.

The numerical procedures, similar to those used in Lee et al. \cite{Lee2015}, are to use particle-based
SRD to replace most parts of Navier-Stokes equation. The cell-wise bulk velocity is further balanced by
the influences from director field and external electric field, which are the last two terms in
Eq.(\ref{Navier-Stokes}). The nematohydrodynamic rotational dynamics described in Eq.(\ref{EL-director})
is represented by the director rotation of SRD particles. The update of the particle directors is due to the
molecular field, shown in Eq.(\ref{LL-potential}) and the shear flow felt by the particle. Because mesoscopic
timescale is longer than the microscopic molecular equipartition timescale, thermal noise $\vec \xi(t)$
controls the velocity distribution of particle rotation.  Therefore the particle-based rotational rule
for SRD directors is 

\begin{equation}
\vec d(n+1)=\vec d(n)+[\vec d(n)\cdot\nabla\vec v - \gamma f(\vec d)]dt+\vec\xi(t) dt.
\label{Langivin}
\end{equation}

This is an overdamped Langevin equation in discrete form, for which the potential well is now a complicated
combination of molecular field, external electric field and the flow gradient field.
This director update rule is similar to the direct angular momentum balance used by Shendruk and
Yeomans \cite{Shendruk2015}, and the angular momentum gain from external field is balanced by the last
term in Eq.(\ref{Navier-Stokes}).

Different nematic liquid crystals exhibit different dielectric properties, when they are subjected to external
electric field. Positive LCs ($\epsilon_{\parallel}>\epsilon_{\perp}$) are polarized along the molecule long-axis,
therefore they align parallel to the electric field. On the other hand, negative LCs
($\epsilon_{\parallel}<\epsilon_{\perp}$) are polarized along the short-axis, leading to perpendicular alignment.
The model clearly shows this effect in the molecular potential $U_{i}$. When a positive LC is considered
($\epsilon_{a}=\epsilon_{\parallel}-\epsilon_{\perp}>0$) the potential minimums appear when
$\vec d_i \parallel \vec E$, and when a negative LC is considered, the minimums appear when
$\vec d_i \perp \vec E$.
To compare our simulation with experiments, we adopt the setups that is commonly considered in EHC study.
Two parallel planar anchoring plates are embracing nematic LCs with a small separation. An alternating electric
potential is applied on these two plates, generating an AC cross-plate electric field.

The simulations are prepared with the following parameters: the rotation angle for SRD collision is
$\alpha_{SRD}=120^o$. The mass and moment of inertial for SRD particles are $m_i = 1$ and $I_i=27$.
This mass-inertial ratio represents an elongated molecule shape for nematic liquid crystals. The average
number of SRD particle per cell is $<N_{Ci}> = 70$ to avoid unrealistic high thermal noise level.
The normalized grid size $ds = 1$ and time step $dt = 1$ are used, while the 3D simulation domain
(Lx,Ly,Lz) = (7,68,28) represents a thin slab geometry. To represent the liquid behavior in the SRD 
algorithm, super-cell collision \cite{Ihle2006} is used to keep finite compressibility small. Spatially 
uniform AC electric field is applied in the cross-plate direction $\vec E_x (t) = E_{x0} \ cos(\omega t)$. 
The amplitude $E_{x0}$ and modulating frequency $\omega$ are the primary free parameters for the EHC pattern 
generation. Some other LC material parameters are the relaxation constant $\gamma = 0.04$, the SRD rotation 
angle $\theta = 120^o$ and the average particle thermal velocity in each degree of freedom 
$v_{th,s} = 0.3 \ ds/dt$, where $s\in (x,y,z,\theta$).

\begin{figure}[h]
\centering
  \includegraphics[height=4.5cm]{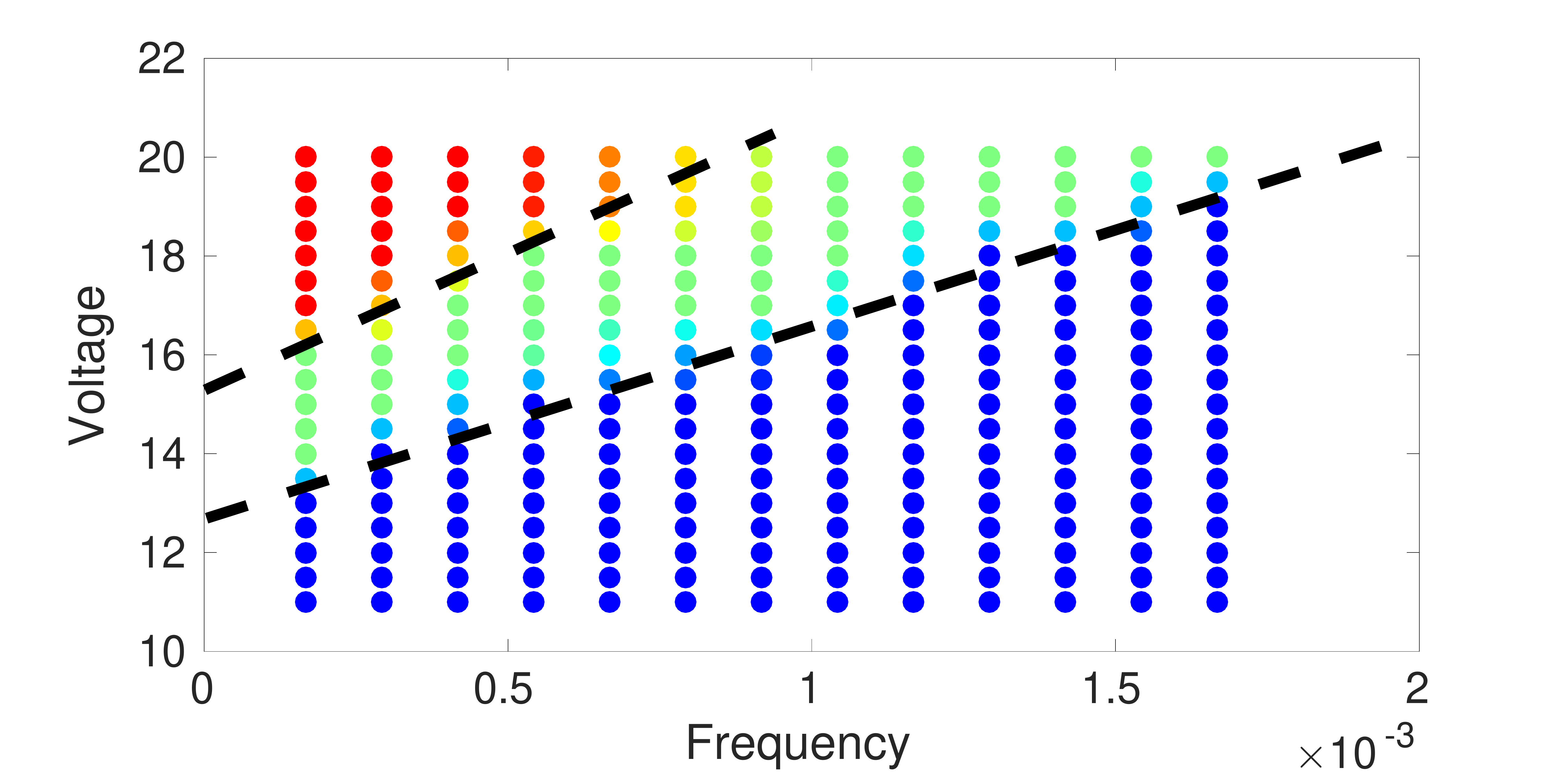}
  \caption{Phase diagram of negative LCs in the normalized voltage and frequency. Blue,
  green and red dots correspond to planar, stationary Williams and fluctuating Williams domains,
  respectively \cite{Sasa1990}.}
  \label{VF_diagram}
\end{figure}

\section{Simulation results and discussion}
The EHC patterns have been mostly studied by using the negative type LC such as MBBA. This is due to that fact that
the weakly nonlinear theory predicts a positive type LC such as 5CB would exhibit more complex nonlinear
behavior \cite{Kramer1995} and it has not been worked out so far, despite the fact that positive LC is widely used
in display experiments and industry.

\begin{figure*}
\centering
  \includegraphics[height=8.5 cm]{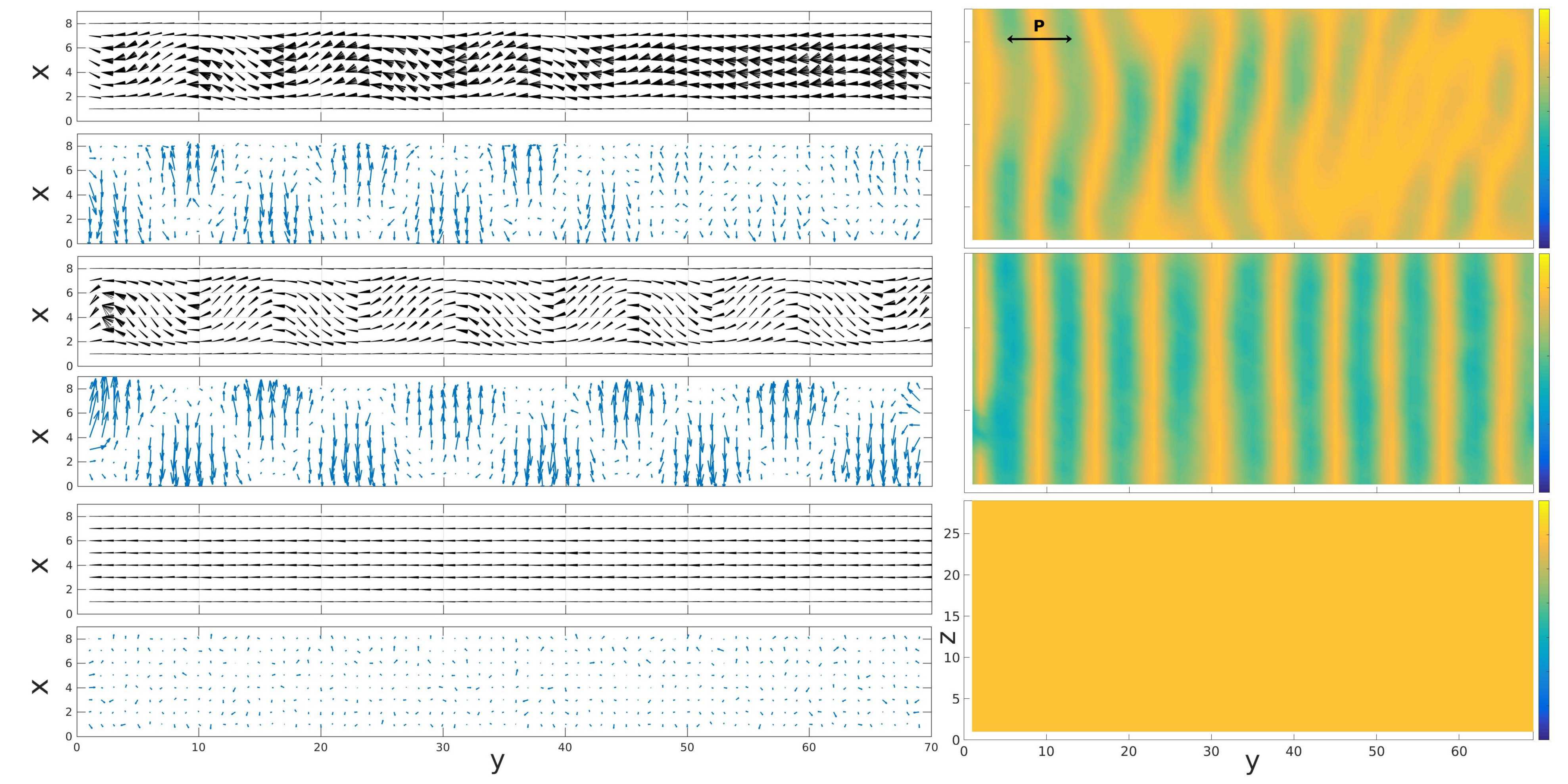}
  \caption{Simulation results of fluctuating Williams, stationary Williams and planar regimes are shown from top to
  bottom rows. In the left column it shows the projected nematic directors (black bars) and the corresponding flow
  field averaged in z-axis (blue arrows). The simulated reflections of linear polarized light (polarization is
  indicated as double-head arrow) are shown in the right column. The simulated frequency is $f = 6.5 \ast 10^{-4}$ and
  the voltages are $dV_{x} = 20$ (fluctuating Williams), $dV_{x} = 17$ (stationary Williams) and $dV_{x} = 10$ (planar)
  respectively.}
  \label{PlaWillTurb}
\end{figure*}

With the desire of creating the positive and negative types of LC in simulation, we specify the different parallel
and perpendicular dielectric constants to represent different orientation tendency under electric field.
For positive LC the parallel and perpendicular dielectric
constants $\epsilon_{\parallel} = 9$ and $\epsilon_{\perp} = 1$ are used such that $\epsilon_{a} = 9$.
For the negative LC we use $\epsilon_{\parallel} = 1$ and $\epsilon_{\perp} = 9$ and the corrsponding
dielectric anisotropy $\epsilon_{a} = 1/9$.

The EHC pattern formation is usually studied in the voltage-frequency phase-space. It is discovered
that, for a fixed AC frequency and increasing voltage, a phase transition appears from planar reflection
to inhomogeneous reflection of the incident polarized light \cite{Williams1963, Kapustin1963}.
By using negative LC this tendency is numerically revealed in our simulations (as shown in Fig.\ref{VF_diagram}).
The originally unperturbed LC directors start to develop regular EHC pattern as the driving voltage increases. As the
voltage increases further the regular EHC patterns evolved into oblique convection rolls, and it is called turbulent state.

Interestingly, we know that the negative LC directors tend to align perpendicularly to the applied electric field,
therefore intuitively one would assume the external field has a stabilization effect on negative LCs. However it
is shown here that this external
field does not further stabilize the system but only introduce free energy to drive the system away from equilibrium. It is
because the linear momentum compensation, the last term in Eq.(\ref{Navier-Stokes}), has a non-zero value
if the charge density appears, hence it works as a source term and the Navier-Stokes equation becomes reaction-diffusion
type. This type of system has chaotic behavior when the driving source term surpasses the diffusion. On the other hand
the cause of the instantaneous charge density is due to the thermal distribution of LC directors, so that there is
always a parallel component along the field direction. This system thermal fluctuation is essential in a non-zero
temperature and a particle-based approach exhibits this effect naturally.

It is seen that the simulated tendency corresponds well with the theoretical prediction \cite{Kramer1995}.
The upshift of triggering voltage when increasing the driving frequency is clearly shown in the phase diagram.
The phase transition from planar to stationary Williams domain is a first-order type, similar to typical RBC phase
transition in normal fluids \cite{Swift1977}. The transition from stationary to turbulent Williams domain is observed
to be, however, a second-order type which shows gradual increase of oblique EHC rolls as increasing voltage.

To investigate the EHC behaviors with the commonly used MBBA and the less studied 5CB, which it is claimed to be more
analytically complicated, we intend to use this simplified particle-based model clarify the details.
In the following we compare the similarities and differences of EHC of two types of LC with positive (a 5CB-like) and
negative (a MBBA-like) anisotropic dielectric constants.

\subsection{Negative LC}

Consider a negative LC, the orientation of molecules tend to be perpendicular to the external electric field direction.
This tendency satisfies the original configuration as shown in Fig.\ref{system}. However the angular momentum, generated 
by the torque that pulls particle director in the y-direction, is converted to transnational momentum in the x-direction
and causes bulk flow if the fluid viscosity can not dissipate its gain.
In Fig.\ref{PlaWillTurb} the typical simulation results are shown for the fluctuating Williams, stationary Williams and
planar phase (from top to bottom). The simulation frequency is normalized to the time-step ($1/dt$), and for the cases 
shown in Fig.\ref{PlaWillTurb} the frequency $f = 6.5 \ast 10^{-4}$ is used. The voltage corresponding to these three 
regimes are $dV_{x} = 20$ (fluctuating Williams), $dV_{x} = 17$ (stationary Williams) and $dV_{x} = 10$ (planar) respectively. 

\begin{figure}[h]
\centering
  \includegraphics[height=4.5cm]{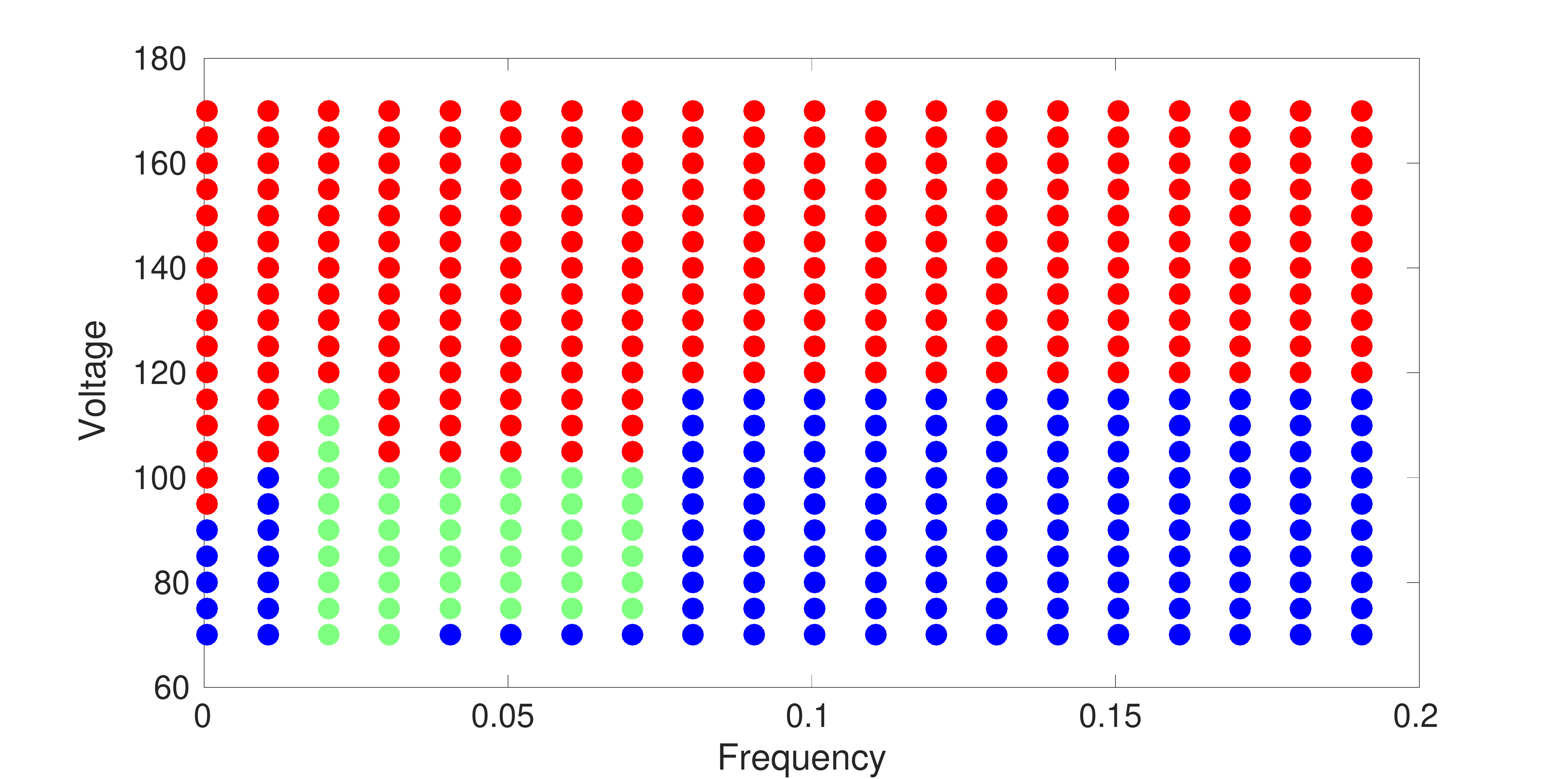}
  \caption{Phase diagram of positive LCs in the normalized voltage and frequency. Blue,
  green and red dots correspond to planar, stationary Williams and fluctuating Williams domains.}
  \label{VF_diagram_5CB}
\end{figure}

In the planar regime the flow caused by external field is suppressed by the fluid viscosity, therefore the velocity 
is about the thermal fluctuation level (as seen in the lowest row of Fig.\ref{PlaWillTurb}). The y-polarized incident 
light has a homogeneous reflection due to the LC complete alignment in y-direction. When the applied electric field
reaches the transition voltage, the collective flow field is suddenly generated and self-organized into convection roll
pattern (as seen in the middle row of Fig.\ref{PlaWillTurb}).  It is observed here that the driving flow only appears
in one of the half planes, depending on the flow direction. Due to the incomprehensibility the flow is strongest in 
the center of driving zone, as an incompressible flow in a nozzle. The flow gradually diverges when it approaches the 
wall. The LC directors exhibit corresponding bending to the flow driving zones, i.e. the upward bending corresponds to
downward driving and vice versa. The patterns of reflected light are periodic in bright-dark stripes, and the separation
of bright-bright stripes (same as dark-dark stripes) is about the channel height $S_{bb,-} \simeq Lx$. This result agrees 
well with the experiment \cite{Sasaki2014}. 

As increasing the driving voltage further the fluid viscosity can no longer hold the regular convection rolls, 
so the system evolves into turbulent Williams regime. To study the system evolution in this regime one should 
take a fully nonlinear approach because the linear/weakly nonlinear assumption, 
i.e. $E = E_{0} + E'$ and $E' \ll E_{0}$ can no longer hold. The most significant signature of turbulent regime 
is the non-steady oblique EHC rolls \cite{Bodenschatz1988}. The director/flow fields, as well as the reflection
pattern, are shown in the top row in Fig.\ref{PlaWillTurb}. It is seen that the zigzag stripes and connected EHC
rolls distribute in the domain, and those features change along time. However, due to our system size and boundary
condition used, the finite-size effect limits the development of those oblique structures. In the much larger 
voltage case, the reflection patterns become patch-like and the flow field is randomized. 

\begin{figure}[h]
\centering
  \includegraphics[height= 4.5 cm]{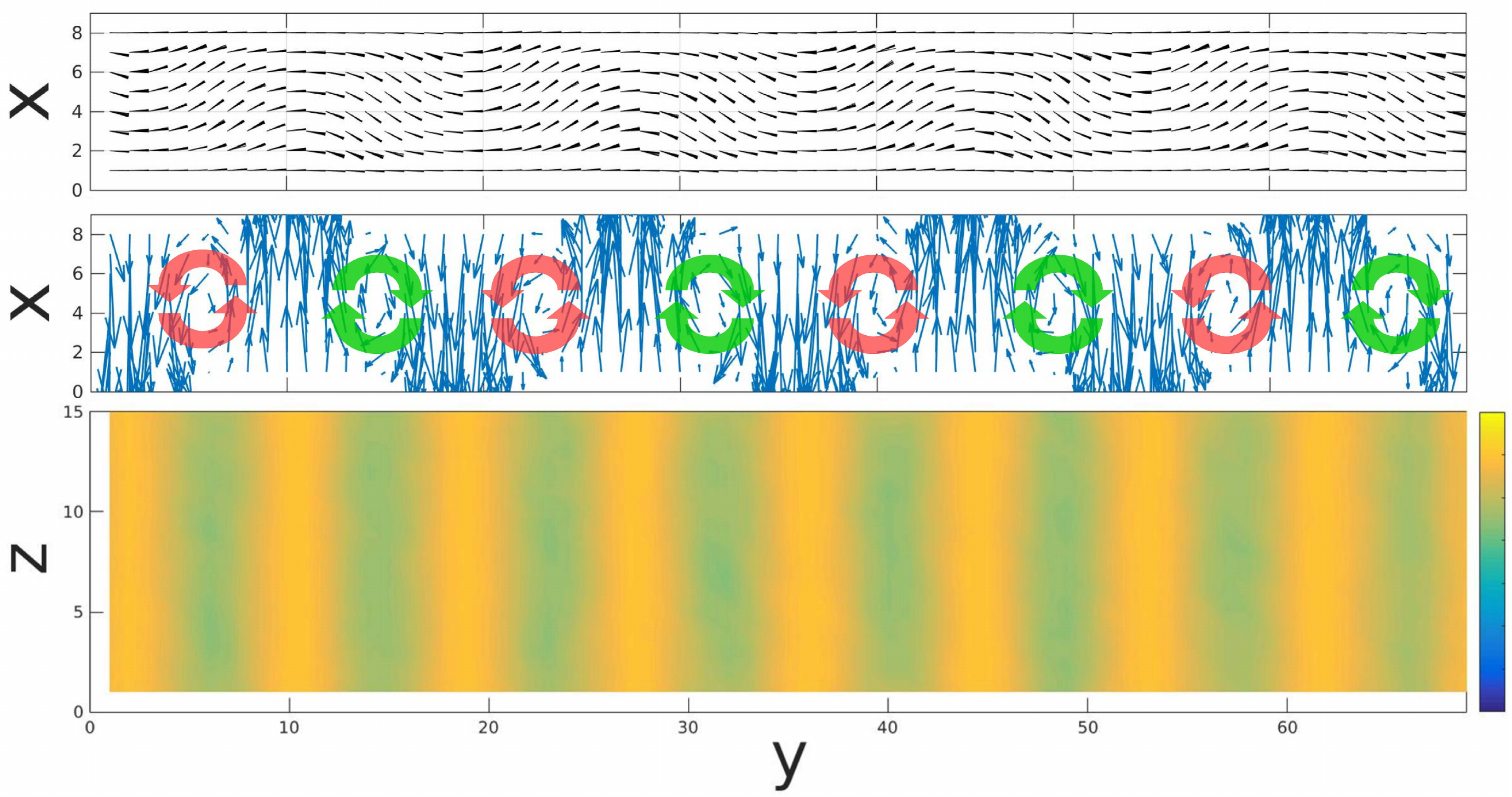}
  \caption{The regular EHC roll of positive LC is shown in its director (top) and flow (middle) fields, viewed from 
  the $z=0$ cross-section. The clockwise green circles and counter-clockwise red circles indicate the flows in the
  convection rolls. The reflection pattern of the y-polarized light is shown in the lowest panel.}
  \label{DVP}
\end{figure}

Detecting the zonal flows is always an important and challenging task for LC microfluidics, due to its scale limitation.
Optical visualization such as doping dye particles in fluid is a common approach. Our simulation reveals the flow 
driving areas of the commonly studied EHC with MBBA molecules. In the case of colloidal particle suspension in EHC, 
such as the nematic colloids transportation \cite{Sasaki2014}, the coupling dynamics can be studied in details.

\subsection{Positive LC}

The more complicated EHC behaviors of positive LC can also be investigated by using positive dielectric anisotropy. 
The particle parameters of simulation are kept the same as those used for negative LC, except the dielectric constant 
for positive LC are now $\epsilon_{\parallel} = 9$ and $\epsilon_{\perp} = 1$, and such that $\epsilon_{a} = 9$. 
As the same procedure we first scan the voltage-frequency phase space for finding different regimes. 

The three different regimes, the turbulent Williams, the Williams and the planar regimes, are all found in 
this VF phase diagram. However the frequency interval of regular Williams EHC appears in the higher frequency
range, and surprisingly the regular rolls do not appear for some frequencies while increasing the driving voltage.
This feature is very different from the results of negative LC, since there the regular EHC always appear between 
planar and turbulent regimes. The DC limit (with vanishing driving frequency $\omega = 0$) as seen in the negative 
LC \cite{Bodenschatz1988} also disappeared for positive LC, the phase transition becomes first order between planar 
and turbulent Williams. 

\begin{figure}[h]
\centering
  \includegraphics[height= 4 cm]{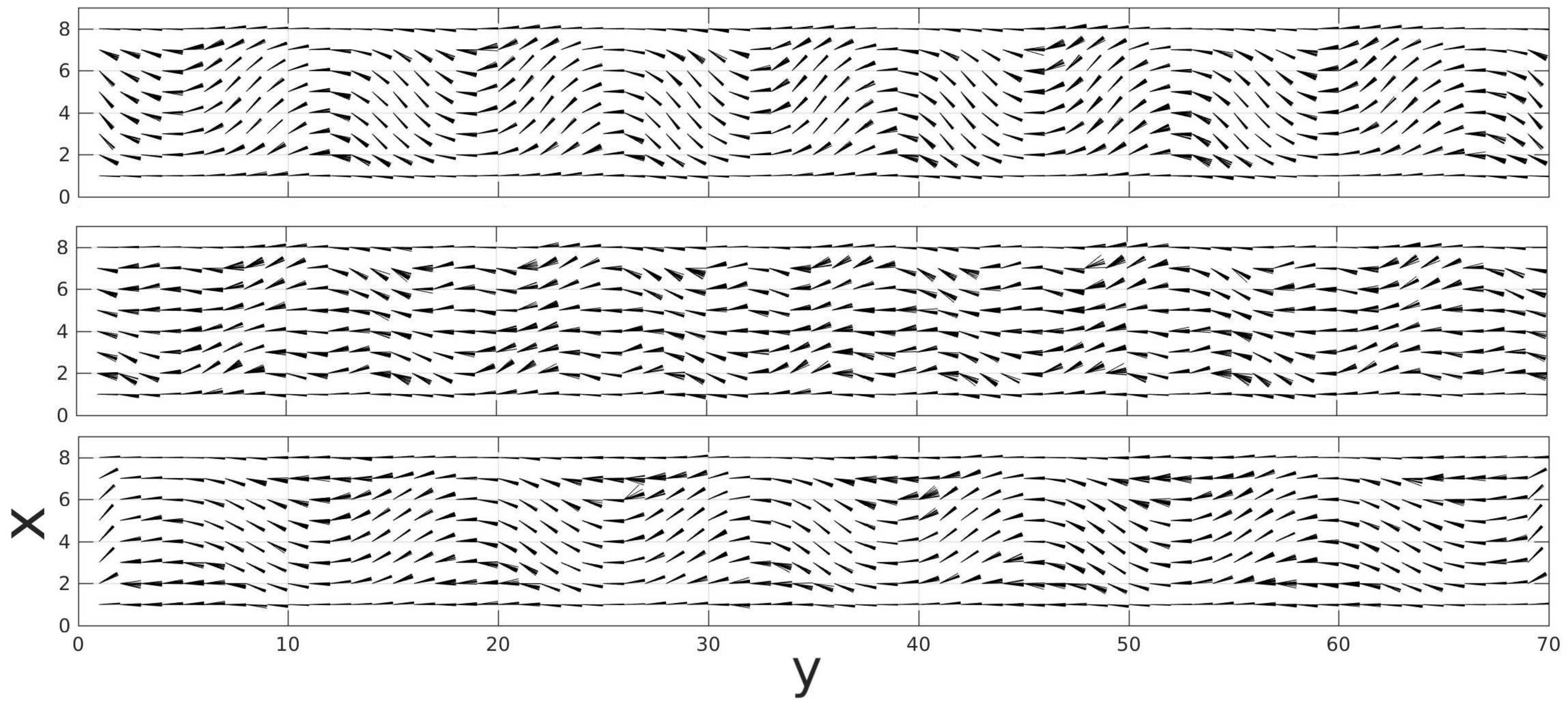}
  \caption{The oscillatory director field of positive LC is shown here. The upper panel is at its positive
  phase, while the middle and the lower panels correspond to the transition and the negative phases.}
  \label{D_evo}
\end{figure}

Another point to notice here is that the starting voltage of EHC rolls is about $dV_{x}\simeq 70$, 
which is higher than that of the negative LC (there is only planar regime below this voltage).
This is not surprising to us because the rotational strength of electric field to the LC particle can always be 
normalized as a number of perfectly aligned particles ($dV_{x}\simeq 70$ has the same alignment strength as including 
$70$ LC neighbors orientating along x-axis around particle $i$), as one can see this analogy in the last two terms in 
Eq.(\ref{LL-potential}).  Qualitatively there are less EHC rolls generated in the same simulation length ($L_{y}$ = 70),
indicating the wider separation of EHC rolls when positive LC is considered. We estimate the difference of stripes 
separation is about $S_{bb,+} \sim 1.2 \ S_{bb,-}$ for comparing positive and negative LCs. This result might be
slightly different in experiment because our simulations are subjected to periodic boundary condition in the y direction.

Another major difference of the regular EHC with positive LC to the one with negative EHC, is that the EHC rolls 
oscillate in time. Figure \ref{D_evo} exhibits this effect in the consecutive time from top to bottom panels. 
It is seen that the region with originally bending upward director changes its configuration and eventually bends
downward after half cycle. Very frequently the occurrence of oscillatory motion indicates the resonance in the system, 
this happens when the frequency of the energy source coincides with the eigen-frequency of the system. The oscillation
frequency of the convection patterns are found to be correlated with the driving frequency, indicating clearly a 
resonance condition. For the case shown in Fig.\ref{D_evo} the electric field oscillation period is $\Delta t = 50$. The 
fully developed EHC rolls reverse the bending direction every $\Delta t = 25$, i.e. two director oscillation cycles is 
a half of the AC field cycle.

\begin{figure}[h]
\centering
  \includegraphics[height= 5 cm]{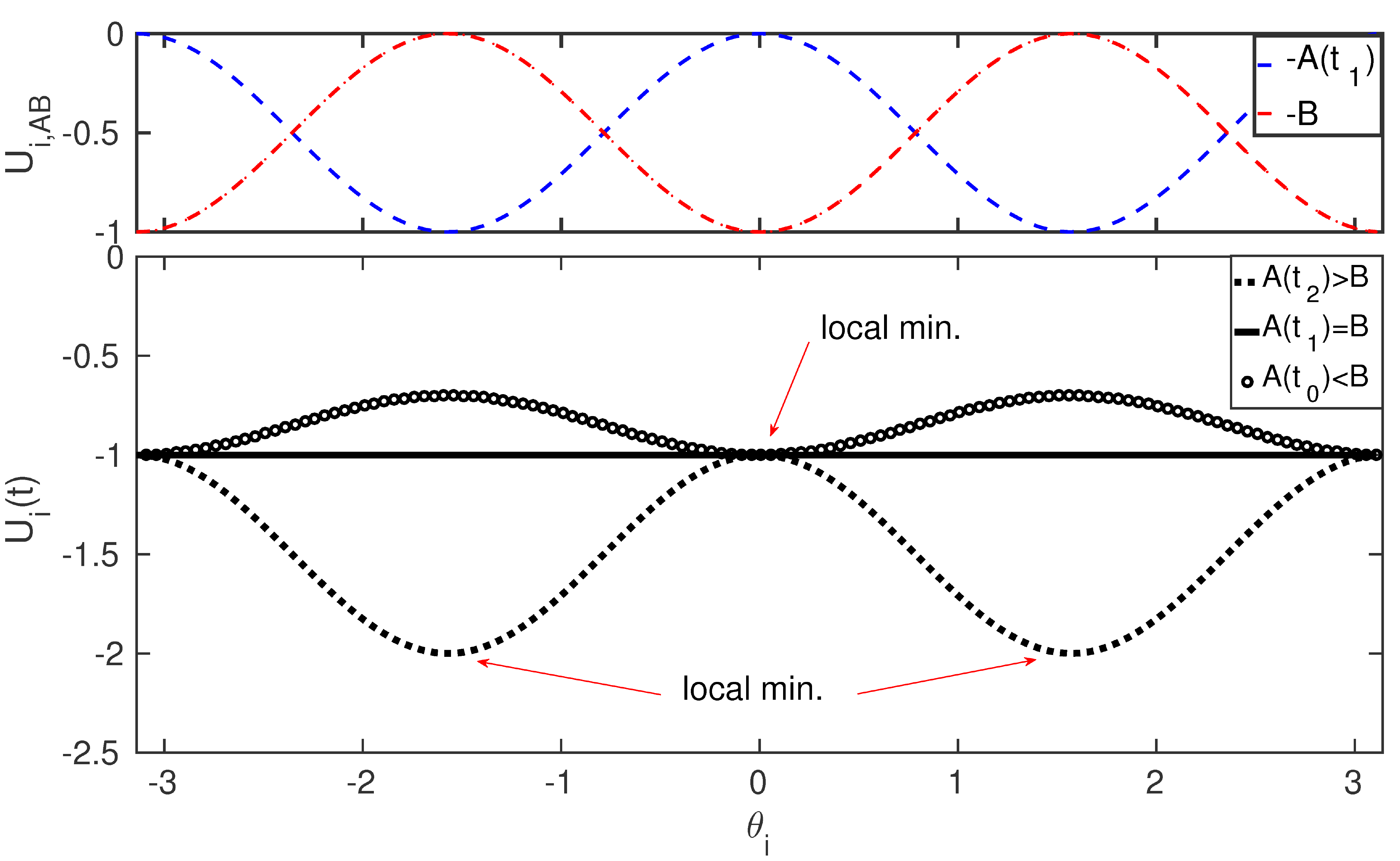}
  \caption{Upper panel shows the contributions from time-dependent $A(t_{1})$ and stationary $B$ to $U_{i,AB}$, as 
  functions of particle angle $\theta_{i}$ in y direction. The lower panel is the sum $U_{i}(t)=-A(t)-B$, 
  when a positive dielectric anisotropy ($\epsilon_{a} > 0$) is assumed. This potential oscillates among
  three phases in time, if the electric field $E_{x0}$ is large enough. The instants $t_{0}$, $t_{1}$ and $t_{2}$
  correspond to $A(t_{0})<B$, $A(t_{1})=B$ and $A(t_{2})>B$, respectively. The local minima of $U_{i}(t)$ are 
  the preferential directions for particles, and they change at different phases of electric field.} 
  \label{oscillate_new}
\end{figure}

This can be simply understood in a molecular potential analysis. Consider a positive LC (i.e. $\epsilon_{a} > 0$), 
the Eq.(\ref{LL-potential}) can be rewritten as
\begin{equation}
U_{i}(t)=-A(t)-B 
\label{potential_new}
\end{equation}
where
\begin{gather}
\nonumber
A(t) = \epsilon_{a} (\vec d_{i} \cdot \vec E_{x0} \ cos(\omega t))^2 \\
\nonumber
B = \sum_{\left\langle i,j\right\rangle} (\vec d_{i} \cdot \vec d_{j})^2
\end{gather}
Here the quantity $A(t)$ is a function 
of time because of the electric field, but $B$ is a stationary function. $A(t)$ and $B$ are out-of-phase in 
particle angle $\theta_{i}$
If the amplitude of electric field $E_{x0}$ is strong enough, i.e. the amplitude of $A(t)$ is 
larger than $B$ for some instants, we then label $t_{0}$, $t_{1}$ and $t_{2}$ for the cases $A(t_{0})<B$, 
$A(t_{1})=B$ and $A(t_{2})>B$.
The individual contributions of $A(t_{1})$ and $B$ are plotted in the upper panel of Fig.\ref{oscillate_new}.
It is noted that the nonlinear feedback to flow is neglected in this potential analysis, so that the 
neighboring $\vec d_{j}$ are always pointing in y direction.

In one cycle of strong electric field, the molecular potential $U_{i}(t)$ can be dominated by either 
$A(t)$ or $B$ at different phase of electric field. These potential profiles are plotted in 
the lower panel of Fig.\ref{oscillate_new}. We see the local minima of particle angle 
$\theta_{i}$ oscillate between $\theta_{i}=0$ and $\theta_{i}=\pi/2$ from time $t_{0}$ to $t_{2}$. This is different 
from a weak electric field situation, which has only one local minimum at $\theta_{i}=0$. This oscillation of potential 
local minima explains why the EHC rolls can resonant with strong electric field.

The oscillatory effect we reported here is also reported in a very recent experiment, where 5CB is 
used as the mesogene \cite{Kumar2010}, although in their work this resonance mechanism was not proposed. 
There the structures of EHC are detected from the intensity of diffraction fringe, and the intensity 
peaks twice in one electric field cycle, which is the same in our simulations with positive LC.

The flow driving zones in the EHC rolls are also different when positive LCs are used. In contrast to the results of 
negative LC, i.e. the driving zones are primarily in the half plane in the flow direction, the driving of positive LC
is in the whole channel height and strongest in the channel center. 

\section{Conclusions}

We have investigated the electrohydrodynamic convection of positive/negative nematic liquid crystals via the newly 
developed particle-based mesoscopic algorithm. Aiming to provide a complementary research tool that can provide
physical insights that can not be studied easily by experiments and linear theory, our nonlinear simulation reveals
the detailed dynamics of the liquid crystal flows and directors.  

The interesting phase transition and pattern formation in EHC are directly simulated, and this deterministic method
shows clearly the dependence of input energy and the system response. Our results correspond well with experiments, 
either for positive or negative LCs considered. Further more, those details that are difficult to measure can also 
be obtained in simulations, this can benefit the experiment setup when the considered system become more complicated,
such as adding colloid suspension in the convection channel.

One of the interesting finding in our simulations is, although they look similar at the first glance, the bright-dark
reflection stripes of positive and negative LCs are actually very different. The characteristic aspect ratio of the 
convection rolls are different, for which positive LC shows a wider structure. As demonstrated in recent experiment,
the EHC patterns with positive LC (5CB) are actually oscillatory. Here we confirm its characteristics numerically and 
we suspect this effect is due to the resonance of incoming energy, which is the AC electric field, with the system.

Also for the driving flow zones these two type of LCs show very different behaviors. This could be significant when
one further considered immerse nematic colloids and study its transportation processes. 

\section*{Acknowledgment}


\end{document}